\documentclass[conf]{new-aiaa}
\usepackage[utf8]{inputenc}

\usepackage{graphicx}
\usepackage{amsmath}
\usepackage[version=4]{mhchem}
\usepackage{siunitx}

\newcommand{\NumAgents}{N}
\newcommand{\AgentTeam}{\mathcal{A}}
\newcommand{\AgentSingle}[1]{#1}
\newcommand{\AgentBottleneck}{\color{green} \AgentTeam \color{black}}

\newcommand{\CRletter}{C} 
\newcommand{\CriticalResource}[2]{\CRletter^\AgentSingle{{#1}}({#2})}

\newcommand{\BottleneckLetter}{B}

\newcommand{\BottleneckMultiple}[2]{\BottleneckLetter( {#2})}

\newcommand{\Protocol}{P}
\newcommand{\ProtocolWait}{wait}
\newcommand{\ProtocolAct}{go}
\newcommand{\ProtocolAgent}[1]{\Protocol^{\AgentSingle{#1}}}

\newcommand{\TimeNeeded}{\gamma}

\newcommand{\CollisionChecker}{W}

\newcommand{\DMdp}{M}
\newcommand{\DMdpStates}{S}

\newcommand{\DMdpObservations}{\Omega}

\newcommand{\DMdpStateSingle}{s}

\newcommand{\DMdpStatesAgent}[1]{\DMdpStates^{\AgentSingle{{#1}}}}

\newcommand{\DMdpObservationsAgent}[1]
{\DMdpObservations^{\AgentSingle{{#1}}}}

\newcommand{\DMdpStateSingleAgent}[1]{\DMdpStateSingle^{\AgentSingle{{#1}}}}

\newcommand{\ai}{\AgentSingle{i}}

\newcommand{\BlockSetAgent}[2]{X^{#1}(#2)}

\usepackage{amssymb}
\usepackage{pgfplots}
\usepackage{hyperref}
\usepackage[capitalise]{cleveref}
\usepackage{longtable,tabularx}
\newtheorem{definition}{Definition}
\title{Separation Assurance in Urban Air Mobility Systems using Shared Scheduling Protocols}

\author{Surya Murthy\footnote{Graduate Research Assistant, Oden Institute for Computational Engineering and Sciences, surya.murthy@utexas.edu, AIAA Student Member.}, Tyler Ingebrand\footnote{Graduate Research Assistant, Oden Institute for Computational Engineering and Sciences}, Sophia Smith\footnote{Graduate Research Assistant, Oden Institute for Computational Engineering and Sciences}, and Ufuk Topcu\footnote{W.A. “Tex” Moncrief, Jr. Chair in Computational Engineering and Sciences VI, Oden Institute for Computational Engineering and Sciences, AIAA Senior Member}}
\affil{ University of Texas at Austin, Austin, Texas, 78712, United States of America}
\author{Peng Wei\footnote{Associate Professor, Department of Mechanical \& Aerospace Engineering, pwei@gwu.edu, AIAA Associate Fellow}}
\affil{George Washington University, Washington DC, 20052, United States of America}
\author{Natasha Neogi\footnote{Senior Technologist, Assured Intelligent Flight Systems, natasha.a.neogi@nasa.gov, AIAA Associate Fellow}}
\affil{NASA Langley Research Center, Hampton, Virginia, 23666, United States of America}

\begin{document}

\maketitle

\begin{abstract}
Ensuring safe separation between aircraft is a critical challenge in air traffic management, particularly in urban air mobility (UAM) environments where high traffic density and low altitudes require precise control. In these environments, conflicts often arise at the intersections of flight corridors, posing significant risks. We propose a tactical separation approach leveraging shared scheduling protocols, originally designed for Ethernet networks and operating systems, to coordinate access to these intersections. Using a decentralized Markov decision process framework, the proposed approach enables aircraft to autonomously adjust their speed and timing as they navigate these critical areas, maintaining safe separation without a central controller. We evaluate the effectiveness of this approach in simulated UAM scenarios, demonstrating its ability to reduce separation violations to zero while acknowledging trade-offs in flight times as traffic density increases. Additionally, we explore the impact of non-compliant aircraft, showing that while shared scheduling protocols can no longer guarantee safe separation, they still provide significant improvements over systems without scheduling protocols.
\end{abstract}

\section*{Nomenclature}
\subsection*{Terminology}
{\renewcommand\arraystretch{1.0}
\noindent\begin{longtable*}{@{}l @{\quad=\quad} l@{}}
$N$& Number of agents in the dec-MDP setting. \\
$M$& dec-MDP tuple. \\
$S$& Joint state space. \\
$s \in S$& Joint state. \\
$S^i$& Individual state space for agent $i$. \\
$s^i \in S^i$& Individual state of agent $i$. \\
$A$& Joint action space. \\
$a \in A$& Joint action. \\
$A^i$& Individual action space for agent $i$. \\
$a^i \in A^i$& Individual action for agent $i$. \\
$T$& Transition function. \\
$R$& Reward function. \\
$\Omega$& Joint observation space. \\
$\Omega^i$& Individual observation space for agent $i$. \\
$\omega^i \in \Omega^i$& Individual observation for agent $i$. \\
$O$& Observation function. \\
$\pi^i$& Individual policy for agent $i$. \\ 
$\pi^{i*}$& Optimal policy for agent $i$. \\
$\CriticalResource{i}{\DMdpStateSingleAgent{i}}$& Set of critical resource states for agent $i$ at state $s^i$. \\
$\beta(s^i, s^j)$& The blocking function that determines if agent $i$ at state $s^i$ can block agent $j$ from state $s^j$. \\
$b(s^i, s^j)$& The set of pairwise bottleneck states between agent $i$ and agent $j$ \\
$B(s^1, ..., s^N)$& The set of bottleneck states for all agents given the current joint state of the dec-MDP. \\
${s^i}' \in B(s^1, ..., s^N)$& Individual bottleneck state for agent $i$. \\
$\gamma$& Necessary time for an agent to access a bottleneck. \\
$P = P^1 \times .... \times P^N$& A shared scheduling protocol, where $P^i$ is the decision rule for agent $i$. \\
$[t^i, t^i + \TimeNeeded]$ & The turn assigned to agent $i$ by the scheduling protocol $P$. \\
$X^i(s^i)$& The set of agents that block agent $i$ at state $s^i$. \\
\end{longtable*}}

\subsection*{UAM Problem Formulation}
{\renewcommand\arraystretch{1.0}
\noindent\begin{longtable*}{@{}l @{\quad=\quad} l@{}}
$x^i$& Position of aircraft $i$  \\
$\phi$& Flight corridor \\
$U^i$& Route for aircraft $i$ \\
$\phi_k^i$& $k^{\text{th}}$ flight corridor on route $U^i$ for aircraft $i$ \\
$\lambda(\phi_i, \phi_j)$&  The intersection function that determines if flight corridors $\phi_i$ and $\phi_j$ are intersecting. \\
$I$& Intersection of flight corridors \\
$d_{\text{LOS}}$& Loss of separation distance
\end{longtable*}}

\subsection*{Methodology}
{\renewcommand\arraystretch{1.0}
\noindent\begin{longtable*}{@{}l @{\quad=\quad} l@{}}
$\phi^i$& Current flight corridor of aircraft $i$ \\
$U^i$& Route of aircraft $i$ \\
$v^i$& Velocity of aircraft $i$ \\
$\dot{v}^{i}$& Acceleration of aircraft $i$ \\
$\psi^i$& The heading of aircraft $i$ \\
$h^j$& Information of the neighboring aircraft $j$ state \\
$d_{comm}$ & The communication distance between aircraft. \\
$a_{\text{halt}}, a_{\text{min}}, a_{\text{max}}, a_{\text{maintain}}$& Speed change actions available to aircraft. \\
$W^i$& Random waiting time used by aircraft $i$ during CSMA/CD. \\
$d^i_{\text{center}}$& Distance from aircraft $i$ to the center of an intersection used by SRTF. \\
$\phi^*$& Priority route defined by Round Robin. \\
\end{longtable*}}

\section{Introduction}

A critical challenge in air traffic management is ensuring safe separation between aircraft.  
This challenge intensifies in urban air mobility (UAM) environments, where numerous aircraft operate over urban areas at low altitudes, leading to high traffic density.
UAM networks can consist of intersecting flight corridors, which increases the likelihood of conflicts between aircraft at these intersections.
UAM environments require a level of precision that exceeds the capabilities of human air traffic controllers.  
Consequently, incorporating autonomous systems is essential for maintaining safe separation between aircraft.

Existing approaches for UAM separation assurance include strategic separation and tactical separation.
Strategic separation methods manage potential conflicts in advance by modifying schedules and routes to balance airspace demand with capacity \cite{DCB}. However, strategic separation methods lack the flexibility required in dense, dynamic UAM environments.
In contrast, tactical separation focuses on real-time deconfliction, enabling aircraft to make immediate speed or path adjustments to avoid conflicts in shared airspace.
Recent advancements \cite{DENIZ2024100157, multiRL2, zhao2021physics} in reinforcement learning enable agents to learn safe navigation policies through repeated simulations, though such methods often require extensive training and may face challenges in real-world implementation.
In this work, we propose a tactical separation approach that leverages shared scheduling protocols to coordinate access to airspace bottlenecks.

We apply shared scheduling protocols to coordinate aircraft movements in UAM environments. 
Shared scheduling protocols were originally developed for use in early Ethernet networks and operating systems \cite{csma} \cite{parekh2016improved}, but have recently been adapted for multi-agent reinforcement learning \cite{ingebrand2023decentralized}.
Shared scheduling protocols were designed to manage access to shared resources and prevent conflicts that arise when multiple agents access a shared resource simultaneously.
These shared resources form bottlenecks, where conflicts between agents are likely to occur.
Shared scheduling protocols assign turns to different agents, controlling which agents can access bottlenecks at a given time.
Shared scheduling protocols prevent conflicts between agents and enhance overall system performance.

UAM networks consist of aircraft navigating along predefined flight corridors to reach their destinations.  
Bottlenecks naturally form at intersections of these corridors. This increases the risk of separation violations as multiple aircraft attempt to occupy the same airspace simultaneously.
We formally identify bottlenecks within the UAM flight network by modeling the system as a decentralized Markov decision process.  
By observing the bottleneck airspace, shared scheduling protocols issue speed adjustment advisories that coordinate when aircraft pass through intersections.  
We present three protocols, two of which are fully decentralized, enabling aircraft to maintain safe separation without a central controller.

We evaluate the proposed scheduling protocols in simulated UAM scenarios with constant-speed aircraft, demonstrating their effectiveness in maintaining safe separation and reducing separation violations to zero.
We further investigate potential trade-offs, noting that while safety improves, flight times increase with higher traffic density.
Finally, we explore the performance of scheduling protocols in settings with non-compliant aircraft, observing a bounded increase in loss of separation events for compliant aircraft as the number of non-compliant aircraft increases.
\section{Related Works}
\subsection{Safety Assurance}
Conflict detection and resolution (CD\&R) systems have been a significant focus of research due to their critical role in ensuring safety in dynamic airspace.
One notable system is Airborne Collision Avoidance System X (ACAS X), which uses probabilistic modeling to predict future aircraft trajectories and compute optimized avoidance maneuvers in real-time \cite{manfredi2016introduction}.
It offers an improvement over its predecessor, Traffic Collision Avoidance System (TCAS) \cite{harman1989tcas}, by incorporating decision logic based on dynamic risk assessment rather than deterministic rule-based responses.
Detect and AvoID Alerting Logic for Unmanned Systems (DAIDALUS) \cite{munoz2015daidalus} detects areas of conflict by predicting aircraft trajectories and provides maneuvering guidance to pilots to prevent or resolve potential conflicts.
Reachability analysis approaches compute the set of all possible future states of an aircraft and determine whether any conflicts arise based on these future states \cite{tomlin1998conflict}. 
Other CD\&R approaches include trajectory optimization approaches \cite{farley2007automated}, dynamic programming \cite{ding2009dynamic}, and rule-based protocols \cite{hwang2007protocol}.
These techniques offer provable safety guarantees, making them particularly useful for safety-critical applications such as air traffic management.
CD\&R approaches operate over relatively short time horizons (2-3 minutes) to resolve individual conflicts between aircraft.
In contrast, our work considers deconfliction over longer time horizons through the application of shared scheduling protocols. Rather than resolving immediate conflicts, these protocols aim to prevent loss of separation events by proactively coordinating aircraft access to critical airspace bottlenecks. Importantly, shared scheduling can complement CD\&R systems in a layered framework: while scheduling protocols prevent conflicts before they arise, CD\&R systems can resolve any remaining conflicts when loss of separation events occur.

Strategic separation management uses pre-planned decisions, including ground delays implemented by air traffic managers. These approaches balance traffic demand with airspace capacity at key bottlenecks such as airport runways, merging points, and air route intersections. Widely used in current air traffic management, strategic separation approaches underpin traffic management initiatives like Ground Delay Programs \cite{odoni1987flow} and Airspace Flow Programs \cite{libby2005operational}. Demand-capacity balancing techniques use optimization and heuristic methods to schedule flight takeoffs, ensuring that demand matches available capacity at bottlenecks \cite{erzberger2014design}.
However, strategic separation approaches often struggle to adapt to real-time variability in aircraft behavior, particularly in dynamic airspace environments. To address this limitation, strategic separation approaches can be integrated with real-time tactical deconfliction methods \cite{DCB}. Shared scheduling protocols make decisions in real time to prevent conflicts. These protocols have the potential to complement strategic separation approaches by accounting for real-time variance and enhancing operational efficiency.

Unlike strategic separation management, tactical separation approaches aim to resolve conflicts between aircraft in real time by enforcing safe separation. These approaches include sampling-based methods like Monte Carlo Tree Search \cite{wu2022safety} and path planning algorithms like \textit{A}* that identify safe trajectories for multiple aircraft \cite{zhao2021multiple}. While these methods focus on resolving conflicts through trajectory planning, we propose shared scheduling protocols as a tactical separation approach that coordinates access to high-conflict areas in UAM environments, hereafter referred to as bottlenecks. By proactively managing entry into these bottlenecks, shared scheduling protocols prevent conflicts and maintain safe operations.

Recently, significant progress has been made in applying deep reinforcement learning methods for tactical separation assurance.
Reinforcement learning approaches model the UAM environment as a Markov Decision Process (MDP) and train aircraft to avoid conflict with each other using local observations.
Several deep reinforcement learning approaches have been proposed, utilizing long-short term memory networks \cite{DENIZ2024100157}, attention layers \cite{multiRL2}, and physics-based models \cite{zhao2021physics}.
These models have proven effective in resolving conflicts and ensuring safe separation between aircraft in simulated UAM environments.
However, deep reinforcement learning methods require an extensive training period during which safety may be compromised. 
In contrast, shared scheduling protocols do not require training, immediately enforcing safety constraints without training violations.

We consider a UAM flight network similar to \citet{multiRL2}, where aircraft travel along pre-specified routes without deviating by changing flight corridors. 
Aircraft are permitted to adjust their speeds to maintain safe separation from neighboring aircraft. 
A key distinction between our setting and that of \citet{multiRL2} is that we assume aircraft can halt and hover in midair.
This assumption is crucial for applying shared scheduling protocols, as it allows aircraft to wait without occupying bottleneck airspace.
\citet{multiRL2} modeled the UAM network as an MDP with each aircraft as an agent trained to use speed adjustments to enforce safe separation constraints via reinforcement learning.
While we also model the UAM environment as an MDP, we do not employ reinforcement learning to ensure safe separation. Instead, we assume aircraft follow a baseline policy and rely on shared scheduling protocols to maintain safe separation.

\subsection{Shared Scheduling Protocols}


Scheduling protocols are used to coordinate access to shared resources among multiple agents, ensuring that tasks are executed efficiently while minimizing conflicts and maximizing resource utilization.
By allocating resources over time, scheduling protocols play a large role in various applications, including operations research \cite{bunte2009overview, abdalkareem2021healthcare}, and computer operating systems \cite{parekh2016improved, silberschatz2000operating}.
The primary objective of scheduling protocols is to optimize performance metrics such as task completion time, resource utilization, and fairness among agents, which can vary significantly depending on the specific context and requirements.


Within the field of computer science, scheduling plays a critical role in operating systems and distributed computing environments. Scheduling algorithms such as Round Robin, Shortest Remaining Time First (SRTF), and First-Come, First-Served (FCFS) dictate the order and allocation of CPU time to various processes, influencing system performance and responsiveness \cite{silberschatz2000operating}. In Round Robin, each process is given a designated time window to access the CPU, allowing for fair time-sharing among processes \cite{parekh2016improved}. 
The length of time windows in Round Robin must be chosen carefully to ensure that processes have sufficient time to utilize the CPU while ensuring that all processes have the opportunity to access the CPU in a timely fashion \cite{tanenbaum2015modern}. 
In contrast, SRTF minimizes the average waiting time by prioritizing processes with the least remaining execution time
However, SRTF may lead to longer processes being perpetually denied access to the CPU, which can significantly impact overall system efficiency \cite{tanenbaum2015modern}.
FCFS processes tasks in the order they arrive, which can lead to inefficiencies such as the "convoy effect," where short processes wait for a long process to complete, ultimately increasing the average waiting time \cite{silberschatz2000operating}.


Outside of computer science, scheduling protocols are widely applied in operations research to optimize resource allocation and service efficiency. For instance, Round-Robin scheduling has been used in humanitarian logistics to balance resource distribution fairly in disaster response efforts \cite{mishra2017dual}. In public transit systems, such as bus networks, optimization approaches like mixed-integer programming are employed to assign vehicles efficiently to specific routes, improving service reliability and reducing operational costs \cite{bunte2009overview}. In healthcare, scheduling optimization is critical for managing patient admissions, nurse rostering, and operating room allocation \cite{abdalkareem2021healthcare}, with models based on integer programming \cite{santibanez2012operations} and metaheuristic algorithms to handle complex and varying demands \cite{aringhieri2015two}.


More recently, shared scheduling protocols have been applied to multi-agent reinforcement learning (MARL) settings, particularly in decentralized environments where conflicts arise over access to shared resources. 
\citet{ingebrand2023decentralized} define bottlenecks within decentralized Markov decision processes as critical resources that multiple agents need to access but can only be used by a subset at a time. 
To resolve conflicts at these bottlenecks, the authors propose using decentralized scheduling protocols that allow each agent to independently manage resource access based on local observations.
This approach avoids centralized controllers, which are often impractical in MARL due to scalability issues.
By penalizing agents for violating the scheduling protocol, the agents learn to respect resource boundaries and reduce conflicts, improving performance in environments with high agent density.
Building on their approach, we extend the use of shared scheduling protocols by modeling the UAM network as a decentralized Markov decision process.
We define bottlenecks as airspace intersections where flight corridors converge, creating natural points of conflict that mirror resource contention in discrete MARL environments.
We implement scheduling protocols to enable aircraft agents to coordinate access to intersection airspace, with two of the protocols being fully decentralized.
\section{Terminology}
\label{section:term}

In this section, we introduce general terms and concepts that underpin our approach, including decentralized Markov decision processes; critical resource states and bottlenecks; and shared scheduling protocols. These concepts provide a flexible framework for modeling and resolving conflicts in a variety of multi-agent systems. While we present these definitions in a general setting here, \cref{section:methodology} will adapt them to the Urban Air Mobility (UAM) context. This progression allows us to first establish the foundational concepts before demonstrating how these concepts apply to identifying and managing areas of conflict between aircraft in UAM systems. By structuring the discussion this way, we aim to clearly delineate the general framework from its specific application.

\subsection{Decentralized Markov Decision Process}
Multi-agent reinforcement learning (MARL) involves multiple agents interacting within a shared environment, where each agent pursues its own objectives while considering the effects of its actions on both the environment and other agents. Decentralized Markov decision processes (dec-MDPs) model multi-agent environments, where agents make decisions based on their own observations without access to the complete system state \cite{DecMdps}.

\begin{definition}
    A decentralized Markov Decision Process (dec-MDP) for \(N\) agents is a tuple \(M = \langle S, A, T, R, \Omega, O \rangle\), where
    \begin{enumerate}
        \item \(S = (S^{1} \times S^{2} \times \ldots \times S^{N})\) is the joint state space, with \(s^{i} \in S^{i}\) representing the state of agent \(i\),
        \item \(A = (A^{1} \times A^{2} \times \ldots \times A^{N})\) is the joint action space, with \(a^{i} \in A^{i}\) representing the action of agent \(i\),
        \item \(T: S \times A \xrightarrow{} S\) is the transition function, mapping the current state \(s \in S\) and joint action \(a \in A\) to the next state \(s' \in S\),
        \item \(R: S \times A \times S \xrightarrow{} \mathbb{R}^N\) is the reward function, with \(r^i = R(s, a, s')\) representing the reward received by agent \(i\) after transitioning from \(s\) to \(s'\) upon taking action \(a\),
        \item \(\Omega = (\Omega^1 \times \Omega^2 \times \ldots \times \Omega^N)\) is the joint observation space, with \(\omega^i \in \Omega^i\) representing the observation of agent \(i\), and
        \item \(O: S \times A \times S \xrightarrow{} \Omega\) is the observation function, which provides the individual observations \(\omega^i \in \Omega^i\) for each agent after taking the joint action \(a\) in state \(s\) and transitioning to \(s'\).
    \end{enumerate}
\end{definition}
At a given time $t$, agents in the dec-MDP exist in a joint state $s_t \in S$. Each agent receives only its own local observation \(\omega^i_t \in \Omega^i\) and does not have access to the full joint state. Agents select actions based on their individual observations, and the environment transitions to a new state \(s_{t+1} \in S\) based on the joint action \(a \in A\). Each agent follows a policy \(\pi^i: \Omega^i \xrightarrow{} \mathcal{P}(A^i)\), which maps observations to actions. The goal of each agent is to learn an optimal policy \(\pi^{i^*}\) that maximizes its expected future reward. While we assume deterministic transition and observation functions in this work, the framework can be extended to handle stochastic environments.

\begin{definition}[Observation Sequence] \,
A joint observation at time $t$ is $\omega_t \in \Omega$. After a joint action $a \in A$ is taken from joint state $s_t \in S$ to next state $s_{t+1} \in S$, the time step increments, resulting in $\omega_{t+1} = O(s, a, s_{t+1})$. Then, $(\omega_{0}^i, \omega_1^{i}, ..., \omega_{\Delta t}^i) \in (\DMdpObservationsAgent{i} \times ... \times \DMdpObservationsAgent{i})$ (a ${\Delta t}$ number of times) represents a sequence of local decentralized dec-MDP observations of length $\Delta t$ for agent $i$. We will represent $(\DMdpObservationsAgent{i} \times ... \times \DMdpObservationsAgent{i})$ as $(\DMdpObservationsAgent{i})^{\Delta t}$ throughout the rest of the paper.
\label{sequence}  
\end{definition}

\subsection{Critical Resources and Bottlenecks}
In the context of dec-MDPs, critical resources are states that agents must pass through to reach their goal under any optimal policy \cite{ingebrand2023decentralized}.
\begin{definition}[Critical Resource]
\label{def: critical resources}
For an agent $\AgentSingle{i}$ at state $\DMdpStateSingleAgent{i} \in \DMdpStatesAgent{i}$, critical resources $\CriticalResource{i}{\DMdpStateSingleAgent{i}} \subset \DMdpStatesAgent{i}$  is the set of states such that $\DMdpStateSingle \in \CriticalResource{i}{\DMdpStateSingleAgent{i}} $  if under any isolated optimal policy, a rolled-out local trajectory from state $\DMdpStateSingleAgent{i}$ that achieves optimal reward will include state $\DMdpStateSingle$. 
\end{definition}
A rolled-out local trajectory is the set of states that an agent $i$ will visit under a policy starting at state $s^i$.
Critical resource states are important when considering conflict in MARL systems because agents must pass through the states if they are behaving optimally. 
Similarly, if an agent is prevented from accessing its critical resource states, then it cannot achieve optimal reward.

We use the definition of critical resources to define bottlenecks in the dec-MDP environment. 
A bottleneck is a set of critical resource states for an agent that can be used to \textit{block} another agent's critical resources. Blocking occurs when one agent is denied access to critical resource states due to the state of another agent.
\begin{definition}[Blocking]
    The blocking function is defined as $\beta(s^i, s^j) \rightarrow \{True, False\}$. An agent $i$ at state $s^i$ can block agent $j$ from state $s^j$ if $\beta(s^i, s^j) = True$.
\end{definition}
The specific blocking function varies between different environments, and we will define blocking in the context of UAM environments in \cref{section:probform}.
\begin{definition}[Pairwise Bottlenecks]
    \label{def: pair-bottleneck}
    Pairwise bottlenecks between a singular agent $i$ in state $s^i$ and agent $j$ ($j\neq i$) in state $s^j$ are defined as $b(s^i,s^j) = \{({s^i}',{s^j}')| {s^i}' \in C(s^i), {s^j}' \in C(s^j), \text{and } \beta({s^i}',{s^j}') = \text{True} \}$.
\end{definition}
A pairwise bottleneck occurs when one agent's access to a critical resource prevents another agent from accessing its own critical resource. Since agents must access critical resources to behave optimally, pairwise bottlenecks represent states where conflicts are likely to arise. These pairwise bottlenecks can be combined to form the collection of all bottleneck states.
\begin{definition}[Collection of Bottlenecks]
    \label{def: bottleneck}
    For dec-MDP $\DMdp$ at time $t$, agents are at states $s^1_t, s^2_t, ... s^N_t$ with critical resources $\CriticalResource{i}{\DMdpStateSingleAgent{i}_t}$. At time $t$, the collection of all bottleneck states is $\BottleneckMultiple{\AgentBottleneck}{\DMdpStateSingleAgent{1}_t, \ldots,
    \DMdpStateSingleAgent{\NumAgents}_t } = \bigcup_{i,j}(b(s^i_t, s^j_t)), i \neq j$.
\end{definition}
The collection of bottlenecks refers to the set of all pairwise bottlenecks across all agents. From this point onward, when we refer to bottlenecks, we are referring to states within the collection of all bottleneck states otherwise specified.

\subsection{Shared Scheduling Protocols}

A scheduling protocol is an algorithm that coordinates access to a bottleneck among multiple agents, ensuring that conflicts are avoided. The protocol assigns the "go" command to a subset of agents, allowing them to enter the bottleneck, while assigning the "wait" command to the remaining agents, preventing their access until their turn. Agents assigned the "wait" command cannot enter bottleneck states until they are explicitly given the "go" command \cite{ingebrand2023decentralized}.

Shared scheduling protocols use observations over time to determine whether an agent should access a bottleneck. Shared scheduling protocols rely on a necessary time, $\TimeNeeded$, which represents the minimum duration required for an agent to safely pass through the bottleneck. During this time, the scheduling protocol allocates access to the bottleneck for a single agent, preventing any blocking agents from entering. By sequentially granting access in this manner, the protocol ensures that agents can safely navigate the bottleneck without contention.

\begin{definition}[Scheduling Protocol] \,
\label{def: ssp}
Given agent $i$ with bottlenecks $\bigcup_{j}(b(s^i_t, s^j_t)), j \neq i$ and necessary time $\TimeNeeded$,
a scheduling protocol $\Protocol = \ProtocolAgent{1} \times ... \times \ProtocolAgent{\NumAgents}$ is defined as a set of decision rules $\ProtocolAgent{i}: (\DMdpObservationsAgent{i})^t \mapsto \{ \ProtocolAct,  \ProtocolWait \}$. For each agent $\ai$, there exists an interval of time $[t^i, t^i + \TimeNeeded]$ such that for all $t' \in [t^i, t^i + \TimeNeeded],$  
\begin{equation*}  
\ProtocolAgent{j}(\omega^{j}_0 ..., \omega^{j}_{t'}) =  \left \{  
        \begin{array}{ll}  
            \ProtocolAct & \quad \text{if } j = i \\  
            \ProtocolWait & \quad \text{if } j \neq i \text{ and } j \in \BlockSetAgent{i}{\DMdpStateSingleAgent{i}},  
        \end{array}  
    \right.  
\end{equation*}  
where $\BlockSetAgent{i}{\DMdpStateSingleAgent{i}}$ is the set of agents that can block agent $i$.  
\label{SSP}  
\end{definition}  
We note that this definition does not address cases where $j \neq i$ and $j \notin \BlockSetAgent{i}{\DMdpStateSingleAgent{i}}$. In such scenarios, the protocol may allow agents to either "go" or "wait" depending on the specific objectives and constraints of the application.

Shared scheduling protocols satisfying \cref{def: ssp} guarantee that each agent $\ai$ is allocated an exclusive time window $[t^i, t^i + \TimeNeeded]$ to act within the bottleneck, during which all other agents that could potentially block $\ai$ are required to wait. \citet{ingebrand2023decentralized} formalizes the guarantees provided by shared scheduling protocols and proves that any protocol satisfying the given definition resolves conflicts for all agents, provided all agents adhere to the protocol. Since each agent is allocated a dedicated turn to access the bottleneck and no conflicting actions occur during this interval, all agents can sequentially and safely pass through the bottleneck without contention. By ensuring that every agent receives a fair opportunity to act, shared scheduling protocols effectively coordinate bottleneck access across multiple agents.
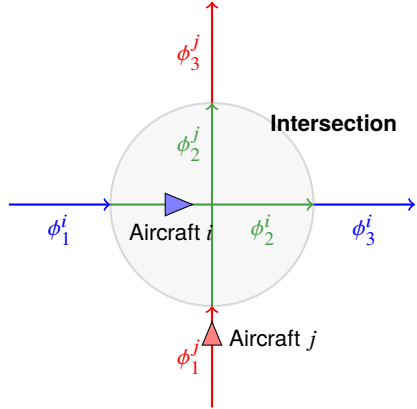
\begin{figure}
    \centering
    \usetikzlibrary{shapes.geometric} 

\begin{tikzpicture}[scale=0.9, every node/.style={font=\sffamily\small}]
    \draw[thick, ->, blue] (-3, 0) -- (-1.5, 0) node[midway, below] {\(\phi^i_1\)};
    \draw[thick, ->, blue] (1.5, 0) -- (3, 0) node[midway, below] {\(\phi^i_3\)};
    \draw[thick, ->, red] (0, -3) -- (0, -1.5) node[midway, left] {\(\phi^j_1\)};
    \draw[thick, ->, red] (0, 1.5) -- (0, 3) node[midway, left] {\(\phi^j_3\)};

    \draw[thick, ->, green!50!black] (-1.5, 0) -- (1.5, 0) node[pos=0.75, below] {\(\phi^i_2\)};
    \draw[thick, ->, green!50!black] (0, -1.5) -- (0, 1.5) node[pos=0.80, left] {\(\phi^j_2\)};

    \draw[thick, gray, fill=gray!20, opacity=0.3] (0, 0) circle (1.5);
    \node at (1.8, 1.2) {\textbf{Intersection}};

    \node[isosceles triangle, draw, fill=blue!50, scale=0.7, shape border rotate=0, label=below:{Aircraft $i$}] (A) at (-0.6, 0) {};
    \node[isosceles triangle, draw, fill=red!50, scale=0.6, shape border rotate=90, label={[xshift=0cm]right:{Aircraft $j$}}] (B) at (0, -2) {};

\end{tikzpicture}
    \caption{Illustration of a UAM flight network with intersecting corridors and aircraft positions. The flight corridors \( \phi^i_1, \phi^i_2, \phi^i_3 \) (blue) and \( \phi^j_1, \phi^j_2, \phi^j_3 \) (red) represent different routes for aircraft \( i \) and \( j \). The intersection, shown as a gray circle with green corridors, occurs where these corridors converge. Aircraft \( i \) and \( j \) are positioned inside and outside the intersection, respectively.}
    \label{fig:intersection}
\end{figure}

\section{Problem Formulation}
\label{section:probform}

In this section, we formalize the Urban Air Mobility (UAM) environment, outlining the assumptions and objectives that underpin our approach. Specifically, we introduce the concept of UAM flight networks, which consist of predefined flight corridors that aircraft must follow to reach their destinations. Within these networks, regions where flight corridors intersect are identified as critical areas where potential conflicts between aircraft may arise. By defining these elements of the UAM environment, we establish the foundation for mapping the general concepts of decentralized Markov decision processes, critical resource states, and bottlenecks to the UAM setting. These definitions play a central role in modeling aircraft behavior and developing strategies to ensure safe separation at intersections.

\subsection{Urban Air Mobility Flight Network}

The Federal Aviation Administration's Concept of Operations for UAM systems introduces UAM flight corridors, designated airspace sections that facilitate point-to-point operations for aircraft \cite{faaconops2023}. These corridors standardize flight paths, enhancing operational efficiency and reducing societal impacts like noise and congestion.

\begin{definition}[Flight Corridor]
    A flight corridor \( \phi \) is a volume of airspace. An aircraft \( i \) at position \( x^i \) is in flight corridor \( \phi \) if \( x^i \in \phi \).
\end{definition}
We assume that flight corridors are \textit{unidirectional}, meaning that aircraft can only fly in one direction through flight corridors. As UAM operations expand, flight corridors can be subdivided and linked to form complex, interconnected flight networks \cite{Environment}. Within these networks, aircraft follow predetermined routes, defined as sequences of flight corridors, to reach their destinations.

\begin{definition}[Routes]
    A route for aircraft \( i \) \( U^i = \{\phi_1^i, \phi_2^i, \ldots, \phi_m^i\} \) is an ordered set of flight corridors \( (\phi_l^i, 1 \leq l \leq m) \).
    \label{Routes}
\end{definition}
For this paper, aircraft flying along route \( U^i \) takes off at the start of \( \phi_1 \) and lands at the end of \( \phi_m \). In our model, aircraft cannot switch routes mid-flight by deviating from their intended flight corridors. Intersections are areas where flight corridors intersect, creating opportunities for interactions among aircraft on different routes.

\begin{definition}[Intersecting Flight Corridors]
    The intersection function \(\lambda(\phi_i, \phi_j)\) is defined as:
    \[
    \lambda(\phi_i, \phi_j) =
    \begin{cases}
    \text{True}, & \text{if } \phi^i \cap\phi^j \neq \emptyset, \\
    \text{False}, & \text{otherwise}.
    \end{cases}
    \]
    Flight corridors $\phi_i$ and $\phi_j$ are intersecting if $\lambda(\phi_i, \phi_j) = \text{True}$.
\end{definition}
By this definition, two flight corridors intersect if their physical airspace overlap (i.e., $\phi^i \cap\phi^j \neq \emptyset$). This means that an aircraft can exist in both flight corridors by flying through the overlapping region. We use the definition of intersecting flight corridors to formally define intersections, which are areas where aircraft on multiple flight corridors can interact.

\begin{definition}[Intersection]
An intersection \(I\) is a set of flight corridors \(\{\phi_1, \phi_2, \ldots, \phi_n\}\) satisfying the following properties:
\begin{enumerate}
    \item \(\forall \phi_i \in I, \; \exists \phi_j \in I, i \neq j \text{ s.t. } \lambda(\phi_i, \phi_j) = \text{True}\), and
    \item \(I\) is closed under intersection: 
    \[
    \phi^* \notin I \implies \lambda(\phi^*, \phi_i) = \text{False}, \; \forall \phi_i \in I.
    \]
\end{enumerate}
An aircraft \(i\) at position $x^{i}$ is within the intersection \(I\) if it is traveling along one of the flight corridors that comprise the intersection. This means \(x^i \in I \iff x^i \in \phi^i_k \; \text{for some } \phi^i_k \in I\).
\label{Intersections}
\end{definition}
An intersection \(I\) is a set of flight corridors that all intersect with at least one other corridor in \(I\). Any corridor that intersects with a member of \(I\) is also included in \(I\). Finally, an aircraft is considered to be in \(I\) if it is traveling along any of the corridors in \(I\).

We illustrate this network in  \cref{fig:intersection}, where the flight corridors \( \phi^i_1, \phi^i_2, \phi^i_3 \) and \( \phi^j_1, \phi^j_2, \phi^j_3 \) represent the routes for aircraft \( i \) and \( j \), respectively The intersection occurs where these corridors converge and aircraft on different routes interact. While \cref{fig:intersection} shows a circular intersection, the shape of the intersection may change depending on the length and position of the intersecting flight corridors. By coordinating when aircraft on different routes pass through intersections, aircraft can maintain a safe separation.

\subsection{Problem Statement}

We consider a UAM flight network where aircraft follow predefined routes that intersect at various points. 
In this model, we assume that aircraft within the same flight corridor can maintain safe separation through speed adjustments. 
This allows us to focus on enforcing safe separation between aircraft traveling on different corridors. 
Additionally, we assume that aircraft fly at a constant altitude and rely solely on speed adjustments to maintain separation. 
This assumption implies that aircraft on the same flight corridor cannot overtake each other, as overtaking would violate the safe separation constraint. 
Furthermore, we assume that aircraft can wait outside intersection airspace in their corridor, which is crucial for applying shared scheduling protocols. 
Without this capability, there would be no way to enforce "wait" commands, rendering the protocols ineffective. 
Our objective is to enforce a minimum safe distance constraint across the network using aircraft speed changes. 
Specifically, we aim to prevent loss of separation (LOS) events, where aircraft come within a specified distance $d_{\text{LOS}}$ of one another.

\section{Methodology}
\label{section:methodology}

In this section, we model the UAM flight network as a decentralized Markov decision process to capture the decision-making dynamics of aircraft navigating the environment. Building on the definitions of flight corridors, routes, and intersections, we formalize the concepts of critical resource states and bottlenecks, which represent key areas of potential conflict under the safe separation objective. By identifying these bottlenecks, we highlight the regions where conflicts between aircraft are most likely to occur. To manage these conflicts, we adapt shared scheduling protocols to the UAM setting, defining rules that coordinate when aircraft can safely traverse bottleneck airspace. This approach integrates the general framework defined in \cref{section:term} with the specific requirements of UAM environments defined in \cref{section:probform}, ensuring safe operations.

\subsection{Flight Network as a Decentralized Markov Decision Process}

We model the UAM flight network as a decentralized Markov decision process (dec-MDP). We represent each aircraft as an agent and define a joint state space, action space, and observation space.
The state space of the UAM dec-MDP is $\DMdpStates = s^{1} \times s^{2}\times ... \times s^{\NumAgents}$ where the $s^{i}$ is the local state of each aircraft. We use an augmented version of the state space defined by \citet{multiRL2}. The state of aircraft $i$ is defined as

\begin{equation}
    s^{i} = (x^{i}, \phi^i, U^i, v^{i}, \dot{v}^{i}, \psi^{i}),
\end{equation}
where, $x^{i}$ is the aircraft's position, $\phi^i$ is the flight corridor, $U^i$ is the route, $v^{i}$ is  velocity, $\dot{v}^i$ is acceleration, and $\psi$ is heading.

Aircraft do not have access to the state information of all other aircraft in the environment. Instead, aircraft observations are comprised of the aircraft's own state information and the state information of neighboring aircraft. The observed state information $h^{j}$ of neighboring aircraft $j$ is

\begin{equation}
    h^{j} = (x^{j}, \phi^j, v^{j}, \dot{v}^{j}, \psi^{j}).
\end{equation}
\noindent The joint observation of the UAM dec-MDP is $\omega \ =  \ \omega^{1} \times \omega^{2}\times ... \times \omega^{\NumAgents}$. The observation of aircraft $i$ is the state information of aircraft $i$ and the states of neighboring aircraft. Formally the observation of aircraft $i$ is
\begin{equation}
    \omega^{i} = s^{i} \cup \bigcup_{j: \|x^i - x^j\| \leq d_{\text{comm}}} h^{j},
\end{equation}
where $d_{\text{comm}}$ is the maximum communication distance for aircraft. To ensure safety between aircraft, the communication distance must be greater than the LOS distance ($d_\text{comm} > d_{\text{LOS}}$). Additionally, we set the communication distance to be greater than the intersection radius. While this isn't required for enforcing safe separation, it provides aircraft with a reasonable view of the intersection before entering.

The action space for each aircraft is comprised of speed change advisories. These include speeding up, slowing down, maintaining current speed, and halting in midair
\begin{equation}
    a = (a_{\text{max}}, a_{\text{min}}, a_{\text{maintain}}, a_{\text{halt}}),
\end{equation}
where
\begin{itemize}
    \item $a_{\text{max}}$ sets the aircraft’s airspeed to $v = 60 kts$,
    \item $a_{\text{min}}$ sets the aircraft’s airspeed to $v = 5 kts$,
    \item $a_{\text{maintain}}$ maintains current speed, and 
    \item $a_{\text{halt}}$ sets the aircraft's airspeed to $v = 0 kts$ and hovers in midair.
\end{itemize}
We assume that the aircraft's minimum and maximum airspeeds are $5 kts$ and $60 kts$ respectively.
Speed changes can be used to increase or decrease the separation between aircraft. 
The $a_{halt}$ action is used by the aircraft to abide by the shared scheduling protocols' "wait" command. 

The aircraft's transition function depends on the aircraft's dynamics. Each aircraft has an existing policy that informs the actions it takes as it navigates the environment. We assume that each agent's policy minimizes the time the aircraft spends halting. This is in line with existing work in aircraft energy consumption, which indicates that halting and hovering is the least desirable state for an aircraft due to its large energy consumption \cite{stahl2018development}.

\subsection{Critical Resources and Bottlenecks in the UAM Flight Network}

For aircraft traversing the flight network, critical resource states correspond to the future flight corridors for the aircraft's route.
Given that aircraft cannot alter their routes mid-flight, all upcoming flight corridors represent states that must be traversed.
Formally, for an aircraft \( i \) at state \(s^{i}\) traveling along flight corridor \( \phi_l^i \) in route \( U^i \), critical resource states are defined as
\begin{equation}
    \CriticalResource{\AgentSingle{i}}{\DMdpStateSingleAgent{i}} = \{s \ | \phi_k^i \in s \forall k \in \{l+1, m\}\}.
\end{equation}
This definition means that every subsequent state along a given route is a critical resource state for aircraft $i$.

To define bottlenecks in the context of the UAM environment, we must first define blocking.
In our UAM setting, we associate blocking with the safe separation constraint, which requires a minimum distance between aircraft.
\begin{equation}
    \beta(s^i, s^j) = 
    \begin{cases}
        \text{True} & \text{if } \| x^i - x^j \| < d_{LOS}, \phi^i \neq \phi^j, v^i \neq 0, \\
        \text{False} & \text{else}.
    \end{cases}
\end{equation}
According to this definition, aircraft can block critical resource states by flying within the LOS distance of the state with non-zero velocity.
Importantly, our definition specifies that blocking only occurs when aircraft are on different flight corridors. 
It is true that an aircraft can prevent aircraft traveling behind along the same route from reaching shared critical resource states. 
However, in such cases, the blocking is not reciprocal, since the aircraft traveling behind cannot overtake and prevent the aircraft traveling in front from reaching the critical resource states.
We can then define the pairwise bottleneck states for aircraft \(i\) and \(j\) as
\begin{equation}
    b(s^i, s^j) = \{({s^i}', {s^j}') | {s^i}' \in \CriticalResource{\AgentSingle{i}}{\DMdpStateSingleAgent{i}}, {s^j}' \in \CriticalResource{\AgentSingle{j}}{\DMdpStateSingleAgent{j}}, \text{and } \beta({s^i}', {s^j}') = \text{ True }\}.
\end{equation}
We recognize that this definition of pairwise bottlenecks includes all areas where different flight corridors intersect and are, consequently, subsets of intersection airspace.
Therefore, instead of coordinating access to sets of states for each agent, we consider the coordination of access to intersections.
An aircraft $i$ is "accessing" an intersection bottleneck $I$ if it is flying through a flight corridor, $\phi^i \in I$, that comprises the intersection at a non-zero airspeed $v^{i} \neq 0$.
We note that in scenarios where no aircraft are traveling along different flight corridors at an intersection, the intersection is no longer a bottleneck. 
However, aircraft would require global information to make that determination. 
Since aircraft only have access to local observations, intersections are always considered bottlenecks from the perspective of each aircraft.

We apply shared scheduling protocols to coordinate access to intersection airspace. 
To improve traffic flow, the UAM scheduling protocols group together aircraft on the same flight corridor into non-conflicting groups, allowing aircraft on the same flight corridor to access the intersection bottleneck at the same time. 
Aircraft receiving the "go" command from the protocol may pass through intersections freely, while those given the "wait" command must halt using the $a_{halt}$ action and wait until they receive the "go" command.

\subsection{Carrier-Sense Multiple Access with Collision Detection}
Carrier-sense multiple access with collision detection (CSMA/CD) is a decentralized scheduling protocol that operates on a first-come first-served basis. 
If an agent is already accessing a bottleneck, CSMA/CD dictates all conflicting agents should wait.
In the UAM setting, this means that if an aircraft is flying through an intersection, all aircraft on different flight corridors should wait.
CSMA/CD \textit{collisions} occur when aircraft on adjacent corridors enter the intersection at the same time. 
These are not the same as actual collisions between aircraft but rather represent \textit{simultaneous access conflicts} that result in delays as aircraft must resolve the contention before proceeding.
In such situations, CSMA/CD initializes a random wait time $\CollisionChecker^i$ for each aircraft. The random waiting times prevent simultaneous access conflicts from repeatedly occurring, as the first aircraft to stop waiting moves through the intersection while the others must keep waiting \cite{csma}.
The maximum waiting time parameter must be chosen carefully, since low values increase the probability of repeated conflict, while large values increase the total waiting time. 
In our simulation, we have aircraft determine their waiting times by uniformly sampling an integer value between $1$ and $100$ simulation time steps, which correspond to $4$ and $400$ seconds, respectively.
One important property of the UAM environment is that aircraft cannot move backward. When simultaneous access conflicts occur, aircraft must exit the intersection by proceeding along their corridor.
Therefore, we decided that aircraft waiting within intersections ($v^{j} = 0$) are not considered to be accessing the bottleneck by the protocols. 
Formally, the CSMA/CD protocol for aircraft $i$ is
 \begin{equation*}
\ProtocolAgent{i}(\omega_0^i ... \omega_{t}^i) =  \left \{
        \begin{array}{ll}
            \ProtocolWait & \quad \text{if } \CollisionChecker^i > 0  \\
            \ProtocolWait & \quad \text{if } \CollisionChecker^i = 0,
             \phi^{i} \notin I,  \\ & \quad \quad \text{and } \exists j : \phi^{j} \in 
             I, \phi^{j} \notin U^{i} \text{ and } v^{j} \neq 0  \\
            \ProtocolWait & \quad \text{if } \CollisionChecker^i = 0,
             \phi^{i} \in I, \text{ and } v^{i} = 0  \\ & \quad \quad \text{and } \exists j : \phi^{j} \in 
             I, \phi^{j} \notin U^{i} \text{ and } v^{j} \neq 0  \\
            \ProtocolAct & \quad \text{else.}
        \end{array}
    \right.
\end{equation*}
The CSMA/CD protocol is defined by several cases that determine whether an aircraft $i$ should act or wait based on the current state of the intersection and the activity of other aircraft. 
First, if the randomized waiting time $\CollisionChecker^i$ is greater than zero, the aircraft must wait to reduce the likelihood of repeated simultaneous access conflicts (Case 1). 
Second, if the aircraft $i$ is not yet in the intersection ($\phi^i \notin I$), it must wait if another aircraft $j$ is moving through the intersection ($v^j \neq 0$) on a different route from $i$ ($\phi^j \notin U^i$) (Case 2).
Similarly, if $i$ is already in the intersection ($\phi^i \in I$) but is stationary ($v^i = 0$), it must continue to wait if another aircraft $j$ is moving through the intersection ($v^j \neq 0$) on a different route from $i$ ($\phi^j \notin U^i$) (Case 3). 
Finally, the protocol allows the aircraft $i$ to proceed through the intersection only if no other aircraft on different routes are moving through the intersection (Case 4).

\subsection{Shortest Remaining Time First}
Shortest Remaining Time First (SRTF) is a decentralized protocol like CSMA/CD but uses different logic when simultaneous access conflicts occur. 
When a conflict occurs, the agent with the shortest remaining time necessary to pass through the bottleneck is allowed to proceed while the other agents must wait \cite{parekh2016improved}. The UAM SRTF implementation handles simultaneous access conflicts by considering the distance of each aircraft to the center of the intersection $d_\text{center}$. Aircraft positioned closer to the center of the intersection are assumed to exit more quickly, given a constant velocity profile, corresponding to a shorter necessary time to exit the bottleneck. 
Formally, the SRTF protocol for aircraft $i$ is
\begin{equation*}
\ProtocolAgent{i}(\omega_0^i ... \omega_{t}^i) =  \left \{
        \begin{array}{ll}
            \ProtocolWait & \quad \text{if } \phi^{i} \in I, \\ & \quad \quad \text{and }\exists j: \phi^{j} \in I, \phi^{j} \notin U^{i}, d_\text{center}^{j} < d_\text{center}^{i}, v^{j} \neq 0,   \\
            \ProtocolWait & \quad \text{if } \phi^{i} \in I, \\ & \quad \quad \text{and }\exists j: \phi^{j} \in I, \phi^{j} \notin U^{i}, d_\text{center}^{j} = d_\text{center}^{i}, v^{j} \neq 0,   \\
            \ProtocolWait & \quad \text{if }
             \phi^{i} \notin I,  \\ & \quad \quad \text{and } \exists j : \phi^{j} \in 
             I, \phi^{j} \notin U^{i} \text{ and } v^{j} \neq 0  \\
            \ProtocolAct & \quad \text{else.}
        \end{array}
    \right.
\end{equation*}

The SRTF protocol is defined by several cases that determine whether an aircraft $i$ should act or wait based on its position relative to the intersection center and the activity of other aircraft. 
First, if $i$ is within the intersection ($\phi^i \in I$), it must wait if there exists another aircraft $j$ also within the intersection ($\phi^j \in I$) on a different route ($\phi^j \notin U^i$), moving ($v^j \neq 0$), and positioned closer to the center of the intersection ($d_\text{center}^j < d_\text{center}^i$) (Case 1). 
Case 1 resolves simultaneous access conflicts by prioritizing the aircraft with the shortest remaining time to exit the bottleneck.
Secondly, if $d_\text{center}^j = d_\text{center}^i$, both aircraft will wait (Case 2).
In these cases, the protocol can use randomized waiting times, similar to CSMA/CD, to ensure one aircraft proceeds first.
Thirdly, if $i$ is not yet in the intersection ($\phi^i \notin I$), it must wait if there exists another aircraft $j$ already within the intersection ($\phi^j \in I$) on a different route ($\phi^j \notin U^i$) and moving ($v^j \neq 0$) (Case 3). 
This condition mirrors the first-come, first-served logic used in Case 2 of the CSMA/CD protocol. 
Finally, the protocol allows the aircraft $i$ to proceed through the intersection only if no other aircraft on different routes are moving through the intersection, ensuring a conflict-free passage (Case 4).

\subsection{Round Robin}
Round Robin is a turn-based protocol where a subset of agents is allowed to access the bottleneck freely \cite{parekh2016improved}.
Unlike CSMA/CD and SRTF, Round Robin is not a decentralized protocol and requires a central authority to keep track of which subset of agents have access to the bottleneck. 
Round Robin operates by selecting a priority flight corridor $\phi^* \in I$. Aircraft on flight corridor $\phi^*$ can travel through the intersection freely while aircraft on other corridors must wait.
The set of blocking agents is comprised of aircraft on different flight corridors than $\phi^*$. 
Formally, the round-robin protocol is
\begin{equation*}
\ProtocolAgent{i}(\omega^{i}_0 ... \omega^{i}_{t}) =  \left \{
        \begin{array}{ll}
            \ProtocolAct & \quad \text{if } \phi^* \in U^{i}\\
            \ProtocolWait & \quad \text{if } \phi^* \notin U^{i}. \\
            \ProtocolWait & \quad \text{if } \phi^* \in U^{i} \text{ and } \exists j: \phi^* \neq \phi^{j} \text{ and } \phi^{j}\in I
        \end{array}
    \right.
\end{equation*}

The Round Robin protocol operates by assigning priority to a specific flight corridor $\phi^*$, allowing only aircraft on that corridor to pass through the intersection while others must wait. 
The protocol can be divided into three cases. 
First, an aircraft $i$ is allowed to proceed if its flight corridor $\phi^i$ is the priority corridor $\phi^*$ (Case 1). 
Second, aircraft $i$ must wait if its flight corridor $\phi^i$ is not the priority corridor $\phi^*$ (Case 2). 
Finally, if there are aircraft on a different flight corridor still in the intersection when the priority switches to $\phi^*$, new aircraft attempting to enter on $\phi^*$ must wait until all aircraft from the previous priority group have exited the intersection (Case 3). 
This ensures safe coordination when aircraft from the old priority group remain in the bottleneck during a turn switch. 

When adapting the Round Robin scheduling protocol to the UAM setting, we assign centralized controllers to each intersection. The aircraft makes entry requests to the controllers when it wants to enter the intersection. The controllers then assign turns to different sets of requesting aircraft based on its route, assigning "wait" and "go" commands depending on which route has the priority. The Round Robin changes priority groups every $100$ simulation time steps, which corresponds to $400$ seconds.
\section{Experiments}

\begin{figure}
    \centering
    \includegraphics[width=0.75\linewidth]{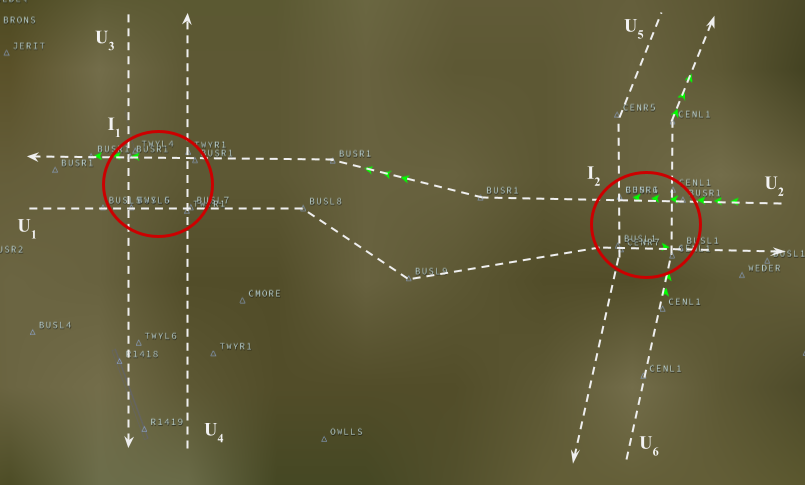}
    \caption{Simulation setting used for tests, consisting of 6 routes and 2 intersections in the BlueSky air traffic simulator. We evaluate shared scheduling protocols in this scenario, varying traffic densities to assess their performance under different conditions.}
    \label{fig:simulation-setting}
\end{figure}

\subsection{Simulation and Baselines}

To test the effectiveness of shared scheduling protocols in urban air mobility (UAM), we use a simulation scenario in the BlueSky Air Traffic Simulator \cite{Simulator}. 
BlueSky is an open-source air traffic management simulator designed to support research with its accessible, modular design and open-data approach, allowing custom scenario creation through text-based commands. 
This flexibility allows us to implement and evaluate conflict detection and resolution algorithms without relying on proprietary data or specialized hardware.
By leveraging these features, we aim to evaluate the potential benefits and limitations of shared scheduling protocols in managing UAM air traffic \footnote{Code for the simulation testing environment is provided at~\url{https://github.com/suryakmurthy/UAM_shared_scheduling/}}.

We define two hyperparameters for the simulation: a Loss of Separation (LOS) distance of \( d_{LOS} = 150\) m and a communication range of \(d_{comm} = 1350\) m. 
The communication distance is equal to the intersection radius, allowing aircraft to detect other aircraft within the intersection before entry, which enables decentralized scheduling protocols to operate effectively. 
Aircraft follow a default policy of constant speed and altitude, traveling at the maximum allowable airspeed ($60$ kts) unless modified by a scheduling protocol. 
To isolate intersection conflicts, we assume that aircraft on the same route can maintain safe separation by adjusting their speed based on the aircraft directly in front of them. 
Aircraft on different routes take off simultaneously to evaluate shared scheduling protocols in settings without strategic separation.

We examine a UAM network of six routes and two intersections to isolate and showcase the specific effects of each scheduling protocol on intersection safety and traffic flow (\cref{fig:simulation-setting}).
We chose this scenario for its simplicity, allowing us to observe the baseline strengths and limitations of scheduling protocols in scenarios where agents have to pass through multiple bottlenecks.
An alternative approach could involve treating pairwise intersections between specific routes as separate intersections.
However, we chose to group such intersections together for simplicity and to ensure adequate spacing between consecutive bottlenecks.
We test four protocols: no protocol (baseline), Carrier Sense Multiple Access with Collision Detection (CSMA/CD), Round Robin, and Shortest Remaining Time First (SRTF).
Each protocol is evaluated across $100$ episodes, during which we record three metrics: LOS events, maximum flight times, and halting times.
Fewer LOS events indicate reduced risk and lower flight times and halting times indicate higher traffic throughput and efficiency.
To further assess scalability, we also vary the number of aircraft on each route to see how the protocols perform under increased traffic densities.

\begin{figure*}[]
    \begin{minipage}{.48\linewidth}
         \begin{tikzpicture}
    
    \begin{axis}[
        title={LOS across different Scheduling Protocols},
        xlabel={Number of Aircraft Per Route},
        ylabel={Average Number of LOS Events},
        xmin=0, xmax=30,
        ymin=-5, ymax=120,
        xtick={5,10,15,20,25},
        ytick={0,20,40,60,80,100},
        legend pos=north west,
        ymajorgrids=true,
        grid style=dashed,
    ]
    
    \addplot[
        color=blue,
        mark=o,
        ]
        coordinates {
        (5,3)(10,31)(15,51)(20,73)(25,106)
        };
    
    \addplot[
        color=red,
        mark=square,
        ]
        coordinates {
        (5,0)(10,0)(15,0)(20,0)(25,0)
        };
    
    \addplot[
        color=purple,
        mark=diamond,
        ]
        coordinates {
        (5,0)(10,0)(15,0)(20,0)(25,0)
        };
        
    \addplot[
        color=orange,
        mark=triangle,
        ]
        coordinates {
        (5,0)(10,0)(15,0)(20,0)(25,0)
        };
        \legend{No Protocol, CSMA/CD, SRTF, Round Robin}
        
    \end{axis}
\end{tikzpicture}
        \caption{Average Number of LOS events in Scenarios with increasing numbers of aircraft per route. All shared scheduling protocols reduce LOS events to 0 even as the number of aircraft increases.}
        \label{LOS_plot}
    \end{minipage}%
    \hfill
    \begin{minipage}{.48\linewidth}
        \begin{tikzpicture}
    
    \begin{axis}[
        title={Maximum Flight times across different Scheduling Protocols},
        xlabel={Number of Aircraft Per Route},
        ylabel={Maximum Flight Times (sec)},
        xmin=0, xmax=30,
        ymin=0, ymax=6000,
        xtick={5,10,15,20,25},
        ytick={0, 1000, 2000,3000,4000,5000, 6000},
        legend pos=south east,
        ymajorgrids=true,
        grid style=dashed,
    ]
    
    \addplot[
        color=blue,
        mark=o,
        ]
        coordinates {
        (5,2944.0)(10,2944.0)(15,2944.0)(20,2944.0)(25,2944.0)
        };
    
    \addplot[
        color=red,
        mark=square,
        ]
        coordinates {
        (5,3226.84)(10,3406.04)(15,3465.72)(20,3612.24)(25,3793.92)
        };
    \addplot[
        color=purple,
        mark=diamond,
        ]
        coordinates {
        (5,3159.96)(10,3324.0)(15,3488.04)(20,3556.96)(25,3784.04)
        };
    \addplot[
        color=orange,
        mark=triangle,
        ]
        coordinates {
        (5,3244.0)(10,3840.04)(15,4168.12)(20,4695.76)(25,5223.64)
        };
        \legend{No Protocol, CSMA/CD, SRTF, Round Robin}
    \end{axis}
\end{tikzpicture}
        \caption{Average Maximum Flight Times in Scenarios with Increasing Numbers of Aircraft. While protocols can reduce LOS events to 0, we note an increase in aircraft flight times.}
        \label{flight_time_plot}
    \end{minipage}%
\end{figure*}

In \cref{LOS_plot}, we show that the introduction of any shared scheduling protocol reduces the number of LOS events to zero. 
This indicates that the protocols effectively coordinate access to bottlenecks and eliminate conflicts between aircraft. 
However, in \cref{flight_time_plot}, we observe that these protocols lead to an increase in aircraft flight times. 
Waiting times at intersections grow as the number of aircraft per route increases, leading to longer travel times.
Specifically, CSMA/CD and SRTF protocols lead to a maximum 15\% increase in flight times, while Round Robin results in a 60\% increase in scenarios with 25 aircraft per route. 
These results show that shared scheduling protocols can ensure safe separation between aircraft while maintaining a bounded increase in travel times.

\begin{table}[]
    \centering
    \begin{tabular}{c|c|c|c|c|c|}
         Protocol & 5 Aircraft & 10 Aircraft & 15 Aircraft & 20 Aircraft & 25 Aircraft \\ \hline
         CSMA/CD & 4.020 \% &  4.665 \% & 7.119 \% & 9.385 \% & 11.771 \%  \\
         SRTF & \textbf{2.393 \%} & \textbf{3.669 \%} & \textbf{6.230 \%} & \textbf{8.372 \%} & \textbf{10.820 \%} \\
         Round Robin & 2.785 \% & 10.653 \% & 16.170 \% & 21.936 \% & 25.983 \%
    \end{tabular}
    \caption{Average percentage of flight time spent halting as the number of aircraft per route increases. Halting times increase with traffic density with SRTF having the shortest halting times.}
    \label{tab:halting times}
\end{table}
\subsection{Results}

Further examination of the protocols' impact on flight times is provided in \cref{tab:halting times}, where we examine the percentage of time spent halting as the number of aircraft per route increases. 
Aircraft following CSMA/CD tend to experience larger halting times compared to those following SRTF. 
This occurs because CSMA/CD resolves simultaneous access conflicts by halting both aircraft, resulting in higher halting times.
In contrast, SRTF uses the distance to the center of the intersection as a tiebreaker, resulting in lower halting times. 
Among all protocols, Round Robin results in the highest halting times.
This is due to the protocol’s turn-based nature, which can cause aircraft at the same intersection to repeatedly halt, leading to greater worst-case delays. 
We observe that, in all cases, halting times increase as the number of aircraft per route rises. 
Higher traffic density at intersections causes longer waiting times, suggesting that integrating traffic flow management strategies, such as demand-capacity balancing, could help reduce halting times for all aircraft.

In scenarios with 5 aircraft, CSMA/CD results in higher average halting times compared to Round Robin due to its reliance on random waiting times. When two aircraft arrive at an intersection simultaneously, CSMA/CD requires each to sample a random waiting time between 1 and 100 simulation time steps. Once the waiting time expires, the aircraft may proceed through the intersection if it is clear, leading to halting times that depend on both the duration of the random waiting time and the time it takes for the first aircraft to clear the intersection. In contrast, Round Robin assigns fixed turns of 100 time steps, allowing aircraft from a designated route to pass through the intersection. If no aircraft are waiting on the active route, Round Robin immediately transitions to a route with pending requests, enabling it to complete turns in less than 100 time steps when fewer aircraft are present. However, as traffic density increases, Round Robin requires more turns to allow all aircraft to pass through the intersection, resulting in repeated waiting periods for the same aircraft at the same intersection. Consequently, while Round Robin outperforms CSMA/CD under low traffic conditions, its advantage diminishes as traffic density rises due to increased delays caused by its fixed-turn structure.

\begin{figure*}[]
    \begin{minipage}{.48\linewidth}
        \begin{tikzpicture}

\definecolor{darkpurple25512714}{RGB}{255,127,14}
\definecolor{forestpurple4416044}{RGB}{44,160,44}
\definecolor{lightgray204}{RGB}{204,204,204}
\definecolor{steelred31119180}{RGB}{31,119,180}

\begin{axis}[
legend cell align={left},
legend style={fill opacity=0.8, draw opacity=1, text opacity=1, draw=lightgray204},
tick align=outside,
tick pos=left,
title={LOSs in Non-Compliant Scenarios},
xlabel={Percent Chance that an Aircraft is Non-Compliant},
xmin=-5, xmax=100,
xtick style={color=black},
ylabel={Average LOS Events per Compliant Aircraft},
legend pos=north west,
ymajorgrids=true,
grid style=dashed,
ymin=-0.0995361758852871, ymax=1.25,
ytick style={color=black}
]

\addplot [line width=0.9pt, red, mark=square, mark size=2.5, mark options={solid}]
table {%
0 0
10 0.00345679012345679
20 0.005
30 0.00714285714285714
40 0.00944444444444444
50 0.01
60 0.0108333333333333
70 0.0103703703703704
80 0.00777777777777778
90 0.0122222222222222
};
\addlegendentry{CSMA/CD}
\addplot [line width=0.9pt, purple, mark=diamond, mark size=2.5, mark options={solid}]
table {%
0 0
10 0.0367901234567901
20 0.0730555555555556
30 0.1
40 0.117777777777778
50 0.135111111111111
60 0.164166666666667
70 0.198888888888889
80 0.234444444444445
90 0.35
};
\addlegendentry{SRTF}
\addplot [line width=0.9pt, orange, mark=triangle, mark size=2.5, mark options={solid}]
table {%
0 0
10 0.0418518518518519
20 0.0838888888888889
30 0.128095238095238
40 0.184814814814815
50 0.241111111111111
60 0.321111111111111
70 0.412592592592592
80 0.547222222222222
90 0.801111111111111
};
\addlegendentry{Round Robin}
\path [draw=red, thick]
(axis cs:0,0)
--(axis cs:0,0);

\path [draw=red, thick]
(axis cs:10,-0.0020864060050972)
--(axis cs:10,0.00899998625201078);

\path [draw=red, thick]
(axis cs:20,-0.00166666666666667)
--(axis cs:20,0.0116666666666667);

\path [draw=red, thick]
(axis cs:30,-0.00106673050221317)
--(axis cs:30,0.0153524447879275);

\path [draw=red, thick]
(axis cs:40,-0.0015096936453113)
--(axis cs:40,0.0203985825342002);

\path [draw=red, thick]
(axis cs:50,-0.00232281834045491)
--(axis cs:50,0.0223228183404549);

\path [draw=red, thick]
(axis cs:60,-0.00327329325066069)
--(axis cs:60,0.0249399599173274);

\path [draw=red, thick]
(axis cs:70,-0.00625921801529159)
--(axis cs:70,0.0269999587560323);

\path [draw=red, thick]
(axis cs:80,-0.0130388822168836)
--(axis cs:80,0.0285944377724391);

\path [draw=red, thick]
(axis cs:90,-0.0225433063270267)
--(axis cs:90,0.0469877507714712);

\addplot [line width=0.9pt, red, mark=-, mark size=5, mark options={solid}, only marks]
table {%
0 0
10 -0.0020864060050972
20 -0.00166666666666667
30 -0.00106673050221317
40 -0.0015096936453113
50 -0.00232281834045491
60 -0.00327329325066069
70 -0.00625921801529159
80 -0.0130388822168836
90 -0.0225433063270267
};
\addplot [line width=0.9pt, red, mark=-, mark size=5, mark options={solid}, only marks]
table {%
0 0
10 0.00899998625201078
20 0.0116666666666667
30 0.0153524447879275
40 0.0203985825342002
50 0.0223228183404549
60 0.0249399599173274
70 0.0269999587560323
80 0.0285944377724391
90 0.0469877507714712
};
\path [draw=purple, thick]
(axis cs:0,0)
--(axis cs:0,0);

\path [draw=purple, thick]
(axis cs:10,0.0100264681014301)
--(axis cs:10,0.0635537788121501);

\path [draw=purple, thick]
(axis cs:20,0.0388866306076037)
--(axis cs:20,0.107224480503507);

\path [draw=purple, thick]
(axis cs:30,0.0554706027329983)
--(axis cs:30,0.144529397267002);

\path [draw=purple, thick]
(axis cs:40,0.071037682695074)
--(axis cs:40,0.164517872860482);

\path [draw=purple, thick]
(axis cs:50,0.0741168794469914)
--(axis cs:50,0.196105342775231);

\path [draw=purple, thick]
(axis cs:60,0.085100453595705)
--(axis cs:60,0.243232879737628);

\path [draw=purple, thick]
(axis cs:70,0.110828708693316)
--(axis cs:70,0.286949069084462);

\path [draw=purple, thick]
(axis cs:80,0.0871268652632671)
--(axis cs:80,0.381762023625622);

\path [draw=purple, thick]
(axis cs:90,0.109437387837656)
--(axis cs:90,0.590562612162344);

\addplot [line width=0.9pt, purple, mark=-, mark size=5, mark options={solid}, only marks]
table {%
0 0
10 0.0100264681014301
20 0.0388866306076037
30 0.0554706027329983
40 0.071037682695074
50 0.0741168794469914
60 0.085100453595705
70 0.110828708693316
80 0.0871268652632671
90 0.109437387837656
};
\addplot [line width=0.9pt, purple, mark=-, mark size=5, mark options={solid}, only marks]
table {%
0 0
10 0.0635537788121501
20 0.107224480503507
30 0.144529397267002
40 0.164517872860482
50 0.196105342775231
60 0.243232879737628
70 0.286949069084462
80 0.381762023625622
90 0.590562612162344
};
\path [draw=orange, thick]
(axis cs:0,0)
--(axis cs:0,0);

\path [draw=orange, thick]
(axis cs:10,0.0195000349470613)
--(axis cs:10,0.0642036687566425);

\path [draw=orange, thick]
(axis cs:20,0.049872734411463)
--(axis cs:20,0.117905043366315);

\path [draw=orange, thick]
(axis cs:30,0.0843792694356302)
--(axis cs:30,0.171811206754846);

\path [draw=orange, thick]
(axis cs:40,0.129818867838574)
--(axis cs:40,0.239810761791055);

\path [draw=orange, thick]
(axis cs:50,0.17158893239573)
--(axis cs:50,0.310633289826492);

\path [draw=orange, thick]
(axis cs:60,0.226151730581238)
--(axis cs:60,0.416070491640984);

\path [draw=orange, thick]
(axis cs:70,0.311974475501442)
--(axis cs:70,0.513210709683743);

\path [draw=orange, thick]
(axis cs:80,0.392188374862491)
--(axis cs:80,0.702256069581954);

\path [draw=orange, thick]
(axis cs:90,0.468781139732448)
--(axis cs:90,1.13344108248977);

\addplot [line width=0.9pt, orange, mark=-, mark size=5, mark options={solid}, only marks]
table {%
0 0
10 0.0195000349470613
20 0.049872734411463
30 0.0843792694356302
40 0.129818867838574
50 0.17158893239573
60 0.226151730581238
70 0.311974475501442
80 0.392188374862491
90 0.468781139732448
};
\addplot [line width=0.9pt, orange, mark=-, mark size=5, mark options={solid}, only marks]
table {%
0 0
10 0.0642036687566425
20 0.117905043366315
30 0.171811206754846
40 0.239810761791055
50 0.310633289826492
60 0.416070491640984
70 0.513210709683743
80 0.702256069581954
90 1.13344108248977
};
\end{axis}

\end{tikzpicture}
        \caption{Number of LOS events at intersections between compliant and non-compliant aircraft. These results are normalized by the number of compliant aircraft. CSMA/CD shows the best performance in settings with high percentages of non-compliant aircraft.}
        \label{fig:non_compliant_2}
    \end{minipage}%
    \hfill
    \begin{minipage}{.48\linewidth}
\begin{tikzpicture}

\definecolor{darkpurple25512714}{RGB}{255,127,14}
\definecolor{forestpurple4416044}{RGB}{44,160,44}
\definecolor{lightgray204}{RGB}{204,204,204}
\definecolor{steelred31119180}{RGB}{31,119,180}

\begin{axis}[
legend cell align={left},
legend style={fill opacity=0.8, draw opacity=1, text opacity=1, draw=lightgray204},
tick align=outside,
tick pos=left,
title={LOSs in Non-Compliant Scenarios},
xlabel={Percent Chance that an Aircraft is Non-Compliant},
xmin=-1, xmax=11,
xtick style={color=black},
ylabel={Average LOS Events per Compliant Aircraft},
legend pos=north west,
ymajorgrids=true,
grid style=dashed,
ymin=-0.00995361758852871, ymax=0.075,
ytick style={color=black}
]

\addplot [line width=0.9pt, red, mark=square, mark size=2.5, mark options={solid}]
table {%
0 0
1 0.000224466891133558
2 0.000113378684807256
3 0.00103092783505155
4 0.000810185185185185
5 0.00210526315789474
6 0.00177304964539007
7 0.0028673835125448
8 0.00277777777777778
9 0.00305250305250305
10 0.00345679012345679
};
\addlegendentry{CSMA/CD}
\addplot [line width=0.9pt, purple, mark=diamond, mark size=2.5, mark options={solid}]
table {%
0 0
1 0.0021324354657688
2 0.00793650793650794
3 0.0119129438717068
4 0.0158564814814815
5 0.0188304093567251
6 0.0237588652482269
7 0.0253285543608124
8 0.0286231884057971
9 0.0313797313797314
10 0.0367901234567901
};
\addlegendentry{SRTF}
\addplot [line width=0.9pt, orange, mark=triangle, mark size=2.5, mark options={solid}]
table {%
0 0
1 0.0037037037037037
2 0.0090702947845805
3 0.0139747995418098
4 0.0179398148148148
5 0.0219883040935672
6 0.0277777777777778
7 0.0329749103942652
8 0.0368357487922705
9 0.0412698412698413
10 0.0418518518518519
};
\addlegendentry{Round Robin}
\path [draw=red, thick]
(axis cs:0,0)
--(axis cs:0,0);

\path [draw=red, thick]
(axis cs:1,-0.00134680134680135)
--(axis cs:1,0.00179573512906846);

\path [draw=red, thick]
(axis cs:2,-0.00101472498538166)
--(axis cs:2,0.00124148235499617);

\path [draw=red, thick]
(axis cs:3,-0.0022472137505737)
--(axis cs:3,0.00430906942067679);

\path [draw=red, thick]
(axis cs:4,-0.00214290528291043)
--(axis cs:4,0.0037632756532808);

\path [draw=red, thick]
(axis cs:5,-0.002388157359602)
--(axis cs:5,0.00659868367539147);

\path [draw=red, thick]
(axis cs:6,-0.00276663460275922)
--(axis cs:6,0.00631273389353937);

\path [draw=red, thick]
(axis cs:7,-0.00223516284445941)
--(axis cs:7,0.00796992986954901);

\path [draw=red, thick]
(axis cs:8,-0.00230474043780213)
--(axis cs:8,0.00786029599335768);

\path [draw=red, thick]
(axis cs:9,-0.00223458732469132)
--(axis cs:9,0.00833959342969743);

\path [draw=red, thick]
(axis cs:10,-0.0020864060050972)
--(axis cs:10,0.00899998625201078);

\addplot [line width=0.9pt, red, mark=-, mark size=5, mark options={solid}, only marks]
table {%
0 0
1 -0.00134680134680135
2 -0.00101472498538166
3 -0.0022472137505737
4 -0.00214290528291043
5 -0.002388157359602
6 -0.00276663460275922
7 -0.00223516284445941
8 -0.00230474043780213
9 -0.00223458732469132
10 -0.0020864060050972
};
\addplot [line width=0.9pt, red, mark=-, mark size=5, mark options={solid}, only marks]
table {%
0 0
1 0.00179573512906846
2 0.00124148235499617
3 0.00430906942067679
4 0.0037632756532808
5 0.00659868367539147
6 0.00631273389353937
7 0.00796992986954901
8 0.00786029599335768
9 0.00833959342969743
10 0.00899998625201078
};
\path [draw=purple, thick]
(axis cs:0,0)
--(axis cs:0,0);

\path [draw=purple, thick]
(axis cs:1,-0.0049114986686928)
--(axis cs:1,0.0091763696002304);

\path [draw=purple, thick]
(axis cs:2,-0.00484062094057216)
--(axis cs:2,0.020713636813588);

\path [draw=purple, thick]
(axis cs:3,-0.00310283583796008)
--(axis cs:3,0.0269287235813736);

\path [draw=purple, thick]
(axis cs:4,-0.00104761120705909)
--(axis cs:4,0.0327605741700221);

\path [draw=purple, thick]
(axis cs:5,0.00117811289622653)
--(axis cs:5,0.0364827058172238);

\path [draw=purple, thick]
(axis cs:6,0.0021244302169006)
--(axis cs:6,0.0453933002795533);

\path [draw=purple, thick]
(axis cs:7,0.00496317729733571)
--(axis cs:7,0.0456939314242892);

\path [draw=purple, thick]
(axis cs:8,0.00585788836703242)
--(axis cs:8,0.0513884884445618);

\path [draw=purple, thick]
(axis cs:9,0.00826085433359401)
--(axis cs:9,0.0544986084258687);

\path [draw=purple, thick]
(axis cs:10,0.0100264681014301)
--(axis cs:10,0.0635537788121501);

\addplot [line width=0.9pt, purple, mark=-, mark size=5, mark options={solid}, only marks]
table {%
0 0
1 -0.0049114986686928
2 -0.00484062094057216
3 -0.00310283583796008
4 -0.00104761120705909
5 0.00117811289622653
6 0.0021244302169006
7 0.00496317729733571
8 0.00585788836703242
9 0.00826085433359401
10 0.0100264681014301
};
\addplot [line width=0.9pt, purple, mark=-, mark size=5, mark options={solid}, only marks]
table {%
0 0
1 0.0091763696002304
2 0.020713636813588
3 0.0269287235813736
4 0.0327605741700221
5 0.0364827058172238
6 0.0453933002795533
7 0.0456939314242892
8 0.0513884884445618
9 0.0544986084258687
10 0.0635537788121501
};
\path [draw=orange, thick]
(axis cs:0,0)
--(axis cs:0,0);

\path [draw=orange, thick]
(axis cs:1,-0.00455213240476004)
--(axis cs:1,0.0119595398121674);

\path [draw=orange, thick]
(axis cs:2,-0.00260275526189002)
--(axis cs:2,0.020743344831051);

\path [draw=orange, thick]
(axis cs:3,-0.00010849291116031)
--(axis cs:3,0.02805809199478);

\path [draw=orange, thick]
(axis cs:4,0.00263969436758994)
--(axis cs:4,0.0332399352620397);

\path [draw=orange, thick]
(axis cs:5,0.00493633032177066)
--(axis cs:5,0.0390402778653638);

\path [draw=orange, thick]
(axis cs:6,0.00913497600994384)
--(axis cs:6,0.0464205795456117);

\path [draw=orange, thick]
(axis cs:7,0.0131897332908153)
--(axis cs:7,0.0527600874977151);

\path [draw=orange, thick]
(axis cs:8,0.0148050285664798)
--(axis cs:8,0.0588664690180612);

\path [draw=orange, thick]
(axis cs:9,0.0188366227210443)
--(axis cs:9,0.0637030598186382);

\path [draw=orange, thick]
(axis cs:10,0.0195000349470613)
--(axis cs:10,0.0642036687566425);

\addplot [line width=0.9pt, orange, mark=-, mark size=5, mark options={solid}, only marks]
table {%
0 0
1 -0.00455213240476004
2 -0.00260275526189002
3 -0.00010849291116031
4 0.00263969436758994
5 0.00493633032177066
6 0.00913497600994384
7 0.0131897332908153
8 0.0148050285664798
9 0.0188366227210443
10 0.0195000349470613
};
\addplot [line width=0.9pt, orange, mark=-, mark size=5, mark options={solid}, only marks]
table {%
0 0
1 0.0119595398121674
2 0.020743344831051
3 0.02805809199478
4 0.0332399352620397
5 0.0390402778653638
6 0.0464205795456117
7 0.0527600874977151
8 0.0588664690180612
9 0.0637030598186382
10 0.0642036687566425
};
\end{axis}

\end{tikzpicture}
        \caption{Number of LOS events at intersections between compliant and non-compliant aircraft. Introducing non-compliant aircraft prevents the scheduling protocols from guaranteeing safe separation between aircraft. }
        \label{fig:non_compliant_1}
    \end{minipage}%
\end{figure*}

\begin{table}[]
    \centering
    \begin{tabular}{c|c|c|c|c|}
         Protocol & 10\% & 20\% & 30\% & 40\% \\ \hline
         CSMA/CD & \textbf{0.28} & \textbf{0.36} & \textbf{0.45} & \textbf{0.51} \\
         SRTF & 2.98 & 5.26 & 6.3 & 6.36 \\
         Round Robin & 3.39 & 6.04 & 8.07 & 9.98 \\
         Baseline & 51 & 51 & 51 & 51 \\
    \end{tabular}
    \caption{Average number of LOS events between compliant aircraft. Even in scenarios with non-compliant aircraft, the protocols can still achieve a significant improvement over baseline behavior.}
    \label{tab:non_comp}
\end{table}
\subsection{Non-Compliant Aircraft}

In addition to scenarios where all aircraft follow the scheduling protocols, we also explore scenarios involving non-compliant aircraft that follow the flight corridor structure and the constant speed and altitude behavior, but ignore the ``wait'' commands issued by the scheduling protocol. These aircraft behave as though they have received the ``go'' command from the protocol at all times, even when they have been asked to ``wait''.  We simulate a scenario with 15 aircraft per route (90 aircraft total), where each aircraft has a fixed probability of being non-compliant at takeoff, and we measure LOS events between compliant and non-compliant aircraft at intersections, based on data from 100 simulation episodes. We focus specifically on LOS events between aircraft on different routes at intersections, as these are the conflicts the shared scheduling protocols aim to prevent. In settings without non-compliant aircraft, we assume that aircraft on the same flight corridor can maintain safe separation by observing the aircraft directly in front of them and adjusting their speed to avoid conflicts. However, in non-compliant settings, non-compliant aircraft may overtake compliant aircraft that are halting at intersections, leading to LOS events. Since the protocols were not designed to prevent these LOS events, we exclude conflicts between aircraft on the same route and only consider those between aircraft on different routes. As we demonstrated in the previous results, aircraft that are following the scheduling protocols will not experience LOS events with each other. Therefore, compliant aircraft only experience LOS events with non-compliant aircraft.

In \cref{fig:non_compliant_1}, we observe that even a small number of non-compliant aircraft can compromise the ability of shared scheduling protocols to guarantee safe separation between aircraft. This limitation arises from the assumptions embedded in each protocol: Round Robin assumes that all aircraft will adhere to the assigned turns, while SRTF and CSMA/CD rely on the expectation that aircraft will follow a first-come, first-served order. Consequently, compliant aircraft passing through intersections operate under the assumption that no other aircraft from conflicting routes will enter simultaneously. This fundamental reliance on protocol adherence makes these approaches unable to guarantee safe separation in scenarios with non-compliant aircraft. In \cref{tab:non_comp}, we observe that these protocols outperform the baseline behavior (51 LOS events).

Among the protocols, CSMA/CD performs the best in all scenarios with non-compliant aircraft (see \cref{fig:non_compliant_2}). 
This is largely due to its first-come-first-served nature, which prevents protocol-compliant aircraft from entering occupied intersections. 
This behavior is also shared by SRTF, which outperforms Round Robin for similar reasons. 
CSMA/CD's performance improves further because it handles simultaneous access conflicts effectively when non-compliant aircraft enter an intersection already occupied by compliant aircraft. 
As noted in \cref{section:methodology}, CSMA/CD resolves such conflicts by having aircraft wait for a random amount of time before attempting to access the bottleneck again. 
This ensures that compliant aircraft already in the intersection do not cause LOS events when non-compliant aircraft enter.
However, CSMA/CD cannot prevent LOS events entirely, as it does not account for scenarios where non-compliant aircraft enter intersections at locations very close to compliant aircraft, resulting in a small number of observed LOS events.
\section{Conclusion and Future Work}

In this work, we presented a new approach for enforcing safe separation between aircraft using shared scheduling protocols. We defined bottlenecks in UAM flight networks as areas where flight corridors intersect and applied shared scheduling protocols to coordinate aircraft access to these bottlenecks, ensuring safe separation in real time. Notably, decentralized protocols such as CSMA/CD and SRTF enable safety enforcement using only local aircraft observations. Numerical experiments demonstrate that integrating these protocols effectively enforces safety constraints with a bounded increase in flight times. In scenarios with non-compliant aircraft, the protocols can no longer guarantee zero LOS events, though they still show significant improvement over baseline performance by reducing the frequency of conflicts compared to systems without scheduling protocols.

Future work will expand upon the baseline explored in this study. A key limitation of shared scheduling protocols is the halting required to coordinate access to intersection airspace. In our experiments, we observed that halting times increase with traffic density at intersections. Future work can address this challenge by exploring strategic separation approaches such as demand-capacity balancing, which can reduce traffic density and, consequently, halting times. Furthermore, shared scheduling protocols could be integrated into reinforcement learning models to optimize halting times by enabling aircraft to reduce speed before reaching intersection boundaries. Finally, future research can modify shared scheduling protocols to account for uncertainty in aircraft observations and improve their robustness in dynamic environments.

\section*{Acknowledgments}
This work was partially supported by the National Aeronautics and Space Administration (NASA) System-Wide Safety (SWS) program under project ``In-time Learning-based Aviation Safety Management System,'' via grant 80NSSC21M0087, and the University Leadership Initiative (ULI) program under project ``Autonomous Aerial Cargo Operations at Scale,'' via grant 80NSSC21M071 to the University of Texas at Austin. The authors are grateful to NASA project technical monitors and partners for their support. Any opinions, findings, conclusions, or recommendations expressed in this material are those of the authors and do not necessarily reflect the views of the project sponsor.

\bibliography{sample}

\end{document}